\documentstyle [12pt] {article}

\catcode`@=12
\begin{document}
\thispagestyle{empty}
\begin{flushright} UCRHEP-T316\\September 2001\
\end{flushright}
\vspace{0.5in}
\begin{center}
{\LARGE \bf Muon g--2 and Neutrino Mass in a New\\ Minimal Extension of the 
MSSM\\}
\vspace{1.2in}
{\bf Ernest Ma\\}
\vspace{0.2in}
{\sl Department of Physics, University of California, 
Riverside, CA 92521, USA} 
\vspace{1.2in}
\end{center}
\begin{abstract}\
If the one term $(\hat \nu_e \hat \mu - \hat e \hat \nu_\mu) \hat 
\tau^c$ is added to the MSSM (Minimal Supersymmetric Standard Model) 
superpotential, the recently observed muon g--2 (anomalous magnetic moment) 
excess can be explained very simply by a light $\tilde \nu_e$.  If the soft 
symmetry-breaking terms $\hat \nu_\alpha \hat h_2^0 - \hat l_\alpha \hat 
h_2^+$ are also added, realistic neutrino masses (with bimaximal mixing) are 
generated as well.
\end{abstract}
\vspace{0.1in}
--------------

\noindent Talk given at the 7th Hellenic School and Workshops on Elementary 
Particle Physics, Corfu, Greece (September 2001).

\newpage
\baselineskip 24pt

\section{Lepton numbers in supersymmetry}

The particle content of the Standard Model with two Higgs doublets as 
required in supersymmetry is given by
\begin{eqnarray}
&&L = (\nu, l)_L \sim (1,2,-1/2), ~~~ \bar E = l^c_L \sim (1,1,1), \\ 
&&Q = (u,d)_L \sim (3,2,1/6), ~~~ \bar U = u^c_L \sim (3^*,1,-2/3), ~~~ 
\bar D = d^c_L \sim (3^*,1,1/3), \\ 
&&\Phi_1 = (\phi_1^0,\phi_1^-)_L \sim (1,2,-1/2), ~~~ \Phi_2 = (\phi_2^+, 
\phi_2^0)_L \sim (1,2,1/2),
\end{eqnarray}
under $SU(3) \times SU(2)_L \times U(1)_Y$.  In the Standard Model without 
supersymmetry, the global quantum numbers $L_e, L_\mu, L_\tau$ and $B$ are 
separately conserved automatically.  In the Minimal Supersymmetric Standard 
Model (MSSM), they are conserved by assumption, i.e. by the removal of the 
allowed terms $\lambda LL\bar E$, $\lambda' LQ\bar D$, $\lambda'' \bar U 
\bar D \bar D$, and $\mu' L \Phi_2$.  In that case, the well-known $R$ 
parity, i.e. $R \equiv (-1)^{3B+L+2J}$, is conserved. 

On the other hand, these terms need not all be forbidden.  In particular, 
the really important restriction is only for the product $\lambda' \lambda''$ 
to be very small or zero to prevent rapid proton decay.  Hence a class of 
$R$-parity violating models has been widely discussed in the literature 
which assumes $\lambda'' = 0$, but allows nonzero values of $\lambda$, 
$\lambda'$, and $\mu'$.  This means that $B$ is conserved, but $L$ is not, 
so that there can be neutrino mass and lepton-flavor violation, etc.

Actually, there are 17 well-defined models of lepton numbers in supersymmetry, 
as pointed out already many years ago \cite{mang}.  Consider the following 
sets of terms in the superpotential:
\begin{eqnarray}
W^{(1)} &=& h_i \Phi_1 L_i \bar E_i + h_{ij}^d \Phi_1 Q_i \bar D_j + 
h_{ij}^u \Phi_2 Q_i \bar U_j + \mu_0 \Phi_1 \Phi_2, \\ 
W^{(2)} &=& f_e L_3 L_1 \bar E_1 + f_\mu L_3 L_2 \bar E_2 + \mu_3 L_3 \Phi_2 
+ f_{ij} L_3 Q_i \bar D_j, \\
W^{(3)} &=& f_{e \mu \tau} L_1 L_2 \bar E_3, \\
W^{(4)} &=& f_{e \mu} L_3 L_1 \bar E_2 + f_{\mu e} L_3 L_2 \bar E_1, \\ 
W^{(5)} &=& f'_e L_2 L_1 \bar E_1 + f'_\tau L_2 L_3 \bar E_3 + \mu_2 L_2 
\Phi_2 + f'_{ij} L_2 Q_i \bar D_j.
\end{eqnarray}
Five models can then be defined with lepton numbers for $(e, \mu, \tau)$ as 
shown below.
\begin{eqnarray}
Model~1 &:& W = W^{(1)} + W^{(2)}, ~~~ [(1,0), ~(0,1), ~(0,0)]\\ 
Model~2 &:& W = W^{(1)} + W^{(3)}, ~~~ [(1,0), ~(0,1), ~(1,1)]\\ 
Model~3 &:& W = W^{(1)} + W^{(2)} + W^{(3)}, ~~~ [1, ~-1, ~0]\\ 
Model~4 &:& W = W^{(1)} + W^{(2)} + W^{(4)}, ~~~ [1, ~1, ~0]\\ 
Model~5 &:& W = W^{(1)} + W^{(2)} + W^{(5)}, ~~~ [1, ~0, ~0].
\end{eqnarray}
Models 1 and 2 have two conserved lepton numbers.  Models 3, 4, and 5 have 
one conserved lepton number.  Each model has also 3 permutations, hence 
there are 1 + 5 $\times$ 3 + 1 = 17 models, ranging from the MSSM with 
conserved $R$ parity to the most general $R$ parity violating model which 
conserves $B$.

\section{Muon anomalous magnetic moment}

In the presence of supersymmetric particles, there are certainly additional 
contributions \cite{susyg2} to the muon anomalous magnetic moment \cite{g-2}. 
To obtain an excess $\Delta a_\mu \sim 10^{-9}$, light $\tilde \nu_\mu$, 
$\tilde \mu$, and large $\tan \beta$ are required, thereby restricting the 
MSSM parameter space.  In this talk I will present a new minimal extension 
\cite{admara} of the MSSM based on Model 2 of the last section, such that 
$\Delta a_\mu$ is explained entirely by the single new term we add to the 
superpotential, i.e.
\begin{equation}
\Delta \hat W = h (\hat \nu_e \hat \mu - \hat e \hat \nu_\mu) \hat \tau^c.
\end{equation}

The new interaction terms of the resulting Lagrangian are then given by
\begin{equation}
{\cal L}_{int} = h (\nu_e \mu - e \nu_\mu) \tilde \tau^c + h (\nu_e \tau^c 
\tilde \mu - e \tau^c \tilde \nu_\mu) + h (\mu \tau^c \tilde \nu_e - \nu_\mu 
\tau^c \tilde e) + H.c.
\end{equation}
Hence there are 2 contributions to the muon anomalous magnetic moment from 
$\tilde \nu_e$ and $\tilde \tau^c$ exchange.  They are easily evaluated 
\cite{kkl} and we obtain
\begin{equation}
\Delta a_\mu = {h^2 m_\mu^2 \over 96 \pi^2} \left( {2 \over m^2_{\tilde 
\nu_e}} - {1 \over m^2_{\tilde \tau^c}} \right).
\end{equation}
Similarly,
\begin{eqnarray}
\Delta a_e &=& {h^2 m_e^2 \over 96 \pi^2} \left( {2 \over m^2_{\tilde 
\nu_\mu}} - {1 \over m^2_{\tilde \tau^c}} \right), \\ 
\Delta a_\tau &=& {h^2 m_\tau^2 \over 96 \pi^2} \left( {2 \over m^2_{\tilde 
\nu_e}} +{2 \over m^2_{\tilde \nu_\mu}} - {1 \over m^2_{\tilde e}} - {1 \over 
m^2_{\tilde \mu}} \right).
\end{eqnarray}

Of all the possible effective four-fermion interactions which can be derived 
from Eq.~(15), only two are easily accessible experimentally: $\mu \to e 
\nu_\mu \bar \nu_e$ through $\tilde \tau^c$ exchange \cite{bgh} and 
$e^+ e^- \to \tau^+ \tau^-$ through $\tilde \nu_\mu$ exchange.  For 
simplicity, both $\tilde \tau^c$ and $\tilde \nu_\mu$ may be assumed to be 
heavy, say a few TeV, 
then the coupling $h$ is allowed to be of order unity in Eq.~(15).  To obtain 
$\Delta a_\mu \sim 10^{-9}$ in Eq.~(16) to account for the possible 
discrepancy of the 
experimental value \cite{g-2} with the standard-model expectation \cite{qcd}, 
we need $\tilde \nu_e$ to be relatively light, say around 200 GeV.

\section{Collider signatures}

Since $\tilde \nu_e$ is required to be light, $\tilde e$ must also be light, 
because of the well-known relationship
\begin{equation}
m^2_{\tilde e} = m^2_{\tilde \nu_e} - M_W^2 \cos 2 \beta.
\end{equation}
Now both $\tilde \nu_e$ and $\tilde e$ can be produced by electroweak 
interactions, such as $Z \to \tilde \nu_e^* \tilde \nu_e$ and $W^- \to 
\tilde \nu_e^* \tilde e$.  They must then decay according to Eq.~(15), i.e.
\begin{equation}
\tilde \nu_e \to \mu^+ \tau^-, ~~~ \tilde e \to \bar \nu_\mu \tau^-.
\end{equation}
These are very distinctive signatures and if observed, the two masses may 
be reconstructed and the value of $\beta$ determined by Eq.~(19).

If the MSSM neutralinos $\tilde \chi^0_i$ and charginos $\tilde \chi^+_i$ 
are produced, as decay products of squarks for example, then the decays
\begin{equation}
\tilde \chi^0_i \to \tilde \nu_e \bar \nu_e (\tilde \nu_e^* \nu_e), ~ 
\tilde e e^+ (\tilde e^* e^-), ~~~ \tilde \chi^+_i \to \tilde \nu_e e^+, ~ 
\tilde e^* \nu_e
\end{equation}
are possible.  The subsequent decays of Eq.~(20) would again be indicative 
of our model. In a future muon collider, the process
\begin{equation}
\mu^+ \mu^- \to \tilde \nu_e^* \tilde \nu_e 
\end{equation}
(through $\tau$ exchange) is predicted, by which the $\tilde \nu_e$ decay of 
Eq.~(20) could be studied with precision.

Single production of $\tilde \nu_e$ and $\tilde e$ is also possible in 
an $e^+ e^-$ collider.  There are 4 different final states:  $\tau^+ \mu^- 
\tilde \nu_e$, $\tau^+ \nu_\mu \tilde e$, and their conjugates.  With the 
subsequent decays given by Eq.~(20), the experimental signatures are 
4 charged leptons $(\tau^+ \tau^- \mu^+ \mu^-)$ and 2 charged taus + missing 
energy $(\tau^+ \tau^- \bar \nu_\mu \nu_\mu)$.  The absence of such events 
at LEP up to 207 GeV constrains $h$ and $m_{\tilde \nu_e}$.  Although a 
quantitative analysis is not available at present, we estimate the likely mass 
bound (on the basis that it would be similar to that of single scalar 
leptoquark production) to be around 180 GeV for $h=1$.  With such a large 
mass, we will need a larger $h$ to get $\Delta a_\mu \sim 10^{-9}$, so we 
have chosen $h=2$ and $m_{\tilde \nu_e} = 200$ GeV (which puts $\tilde \nu_e$ 
beyond the production capability of LEP) as representative values.

\section{Neutrino masses}

Our model as it stands forbids neutrino masses because it conserves $L_e$ and 
$L_\mu$ (with $L_\tau = L_e + L_\mu$).  Consider now the \underline {soft} 
breaking of these lepton numbers by the terms
\begin{equation}
\mu_\alpha (\hat l_\alpha \hat h_2^+ - \hat \nu_\alpha \hat h_2^0)
\end{equation}
in the superpotential, i.e. the so-called bilinear $R$-parity violation 
\cite{cf}.  In that case, the $4 \times 4$ neutralino mass matrix of the 
MSSM must be expanded to include the 3 neutrinos as well to form a 
$7 \times 7$ mass matrix.  It is well-known that \underline {one} tree-level 
mass, corresponding to a linear combination of $\nu_e$, $\nu_\mu$, and 
$\nu_\tau$ is now obtained.  In this scenario, the scalar neutrinos also 
acquire nonzero vacuum expectation values \cite{drv} and one-loop radiative 
neutrino masses are possible \cite{brpv}.  To fit the present data on 
atmospheric \cite{atm} and solar \cite{solar} neutrino oscillations, 
restrictions on the parameters of the MSSM are implied.

In our model there is another, unrestricted source of radiative neutrino 
mass, which gives a contribution only to the off-diagonal $\nu_e \nu_\mu$ 
term through $\tilde \tau^c - h_1^-$ mixing.  Hence our effective $3 \times 
3$ neutrino mass matrix in the basis $(\nu_e, \nu_\mu, \nu_\tau)$ is of the 
form
\begin{equation}
{\cal M}_\nu = \left[ \begin{array} {c@{\quad}c@{\quad}c} a_1^2 & a_1 a_2 + b 
& a_1 a_3 \\ a_1 a_2 + b & a_2^2 & a_2 a_3 \\ a_1 a_3 & a_2 a_3 & a_3^2 
\end{array} \right],
\end{equation}
where we have assumed that the usual one-loop contributions from bilinear 
$R$-parity violation \cite{brpv} are actually negligible, which is the case 
for most of the MSSM parameter space.  This matrix has 4 parameters and 
yields 3 eigenvalues and 3 mixing angles. Consider for example $a_3 = a_2$ 
and define $x \equiv 1 + (b/a_1 a_2)$, we then have
\begin{equation}
{\cal M}_\nu = \left[ \begin{array} {c@{\quad}c@{\quad}c} a_1^2 & x a_1 a_2 & 
a_1 a_2 \\ x a_1 a_2 & a_2^2 & a_2^2 \\ a_1 a_2 & a_2^2 & a_2^2 \end{array} 
\right].
\end{equation}
Assuming that $a_1$ and $x a_1$ are much smaller than $a_2$, the eigenvalues 
are easily determined to be
\begin{eqnarray}
m_{1,2} &=& \mp {(1-x) a_1 a_2 \over \sqrt 2} + {(1-x)(3+x) a_1^2 \over 8}, \\ 
m_3 &=& 2 a_2^2 + {(1+x)^2 a_1^2 \over 4},
\end{eqnarray}
corresponding to the eigenstates
\begin{equation}
\left[ \begin{array} {c} \nu_1 \\ \nu_2 \\ \nu_3 \end{array} \right] = 
\left[ \begin{array} {c@{\quad}c@{\quad}c} 1/\sqrt 2 & 1/2 & -1/2 \\ 
1/\sqrt 2 & -1/2 & 1/2 \\ 0 & 1/\sqrt 2 & 1/\sqrt 2 \end{array} \right] 
\left[ \begin{array} {c} \nu_e \\ \nu_\mu \\ \nu_\tau \end{array} \right],
\end{equation}
to order $a_1/a_2$, which is of course very near the case of bimaximal mixing. 
Atmospheric neutrino oscillations are thus explained by $\nu_\mu \to \nu_\tau$ 
with $\sin^2 2 \theta \simeq 1$ and
\begin{equation}
\Delta m_{23}^2 \simeq \Delta m_{13}^2 \simeq 4 a_2^4 + {1 \over 2} 
(1 + 6x + x^2) a_1^2 a_2^2, 
\end{equation}
and solar neutrino oscillations by $\nu_e \to (\nu_\mu - \nu_\tau)/ 
\sqrt 2$ with $\sin^2 2 \theta \simeq 1$ and
\begin{equation} 
\Delta m_{12}^2 \simeq {(1-x)^2 (3+x) \over 2 \sqrt 2} a_1^3 a_2.
\end{equation}
Using $a_2 = 0.16$ eV$^{1/2}$, $a_1 = 0.05$ eV$^{1/2}$, and $x=-1$, we find 
$\Delta m_{atm}^2 \simeq 2.5 \times 10^{-3}$ eV$^2$, and $\Delta m_{sol}^2 
\simeq 5.7 \times 10^{-5}$ eV$^2$, in very good agreement with data.

The parameter $b$, i.e. the radiative $\nu_e \nu_\mu$ mass, is given by
\begin{equation}
b = {G_F m_\mu^2 \over 4 \pi^2 \sqrt 2}  {h A m_\tau 
\langle \tilde \nu_\tau \rangle \over m^2_{eff} \cos^2 \beta},
\end{equation}
where $m_{eff}$ is a function of $m_{\tilde \tau^c}$ and $m_{h^\pm}$. 
Using $h = 2$ and $m_{eff}^2/A = 1$ TeV, we find that in order to  
obtain $b = -2 a_1 a_2 \simeq 0.016$ eV, we need $\langle \tilde \nu_\tau 
\rangle \simeq 1.93 \cos^2 \beta$ GeV.  This relatively small value is 
negligible compared to $v = (2 \sqrt 2 G_F)^{-1/2} = 174$ GeV (especially 
for large values of $\tan \beta$), and consistent with all present 
low-energy phenomenology.

\section{Lepton flavor violation}

Lepton-flavor violating processes are very much suppressed in our model, 
because they have to be proportional to the small parameters $\mu_\alpha$ in 
Eq.~(23) or the small vacuum expectation values $\langle \tilde \nu_\alpha 
\rangle$.  For example, the rare decay $\tau \to e \gamma$ proceeds in 
one-loop order through $\tilde \nu_e$ exchange and the mixing of $\mu_L$ 
with $\tilde w^-$, and through $\tilde e$ exchange and the mixing of 
$\nu_\mu$ with $\tilde B$ and $\tilde w^0$.  This amplitude and the parameter 
$a_2$ in Eq.~(24) are both proportional to
\begin{equation}
\left( {\mu_\mu \over \mu_0} - {\langle \tilde \nu_\mu \rangle \over 
v \cos \beta} \right),
\end{equation}
where $\mu_0$ is the coefficient of the $(\hat h_1^- \hat h_2^+ - \hat h_1^0 
\hat h_2^0)$ term in the superpotential of the MSSM.  Hence
\begin{equation}
{B(\tau \to e \gamma) \over B(\tau \to e \nu \bar \nu)} \propto {a_2^2 M_W^4 
\over m_\tau^2 m_{eff}^3}
\end{equation}
Let $m_{eff} = 200$ GeV, then $B(\tau \to e \gamma) \sim 10^{-13}$, which is 
many orders of magnitude below the experimental upper bound of $2.7 \times 
10^{-6}$.  

The $\mu \to e \gamma$ rate is even more suppressed because it has to 
violate both $L_\mu$ and $L_e$, whereas $\tau \to e \gamma$ only needs to 
violate $L_\mu$.  We note that if we had chosen the extra term in Eq.~(14) 
to be $h(\hat \nu_e \hat \tau - \hat e \hat \nu_\tau) \hat \mu^c$ or 
$h(\hat \nu_\mu \hat \tau - \hat \mu \hat \nu_\tau) \hat e^c$, then $\mu \to 
e \gamma$ would not be doubly suppressed and would have a branching fraction 
of about $4 \times 10^{-10}$, in contradiction with the present experimental 
bound \cite{meg} of $1.2 \times 10^{-11}$.

\section{Conclusion}

In conclusion, we have shown how a novel minimal extension of the MSSM with 
$L_\tau = L_e + L_\mu$ allows it to have a large contribution to the muon 
anomalous magnetic moment without otherwise constraining the usual MSSM 
parameter space.  With the soft and spontaneous breaking of this lepton 
symmetry, realistic neutrino masses (with bimaximal mixing) are generated 
for a natural explanation of atmospheric and solar neutrino oscillations. 
The scalar electron doublet $(\tilde \nu_e, \tilde e)$ is predicted to be 
light (perhaps around 200 GeV) and has distinctive experimental signatures.

\section*{Acknowledgments}

I thank the organizers George Zoupanos, Nick Tracas, and George Koutsoumbas 
for their great hospitality at Corfu.  This work was supported in part by the 
U.~S.~Department of Energy under Grant No.~DE-FG03-94ER40837.  

\bibliographystyle{unsrt}

\end{document}